\documentstyle[preprint,prb,aps,epsf]{revtex}
\begin{document}

\draft

\title{\bf Influence of the biquadratic interlayer coupling in the specific heat of Fibonacci
magnetic multilayers}

\author{C.G. Bezerra$^{\rm a}$, E. L. Albuquerque $^{\rm b}$ \thanks
{Corresponding author,  e-mail: eudenilson@dfte.ufrn.br} and M.G. Cottam$^{\rm a}$}
 \maketitle

\noindent (a) Department of Physics and Astronomy, University of
Western Ontario, London, Ontario N6A 3K7, Canada.

\noindent (b) Departamento de F\'\i sica, Universidade Federal do
Rio Grande do Norte, 59072-970, Natal-RN, Brazil.

\begin{abstract}

A theoretical study of the specific heat C(T) as a function of
temperature in Fibonacci magnetic superlattices is presented. We
consider quasiperiodic structures composed of ferromagnetic films,
each described by the Heisenberg model, with  biquadratic and
bilinear coupling between them. We have taken the ratios between
the biquadratic and bilinear exchange terms according to
experimental data recently measured for different regions of their
regime. Although some previous properties of the spin wave
specific heat are also reproduced here, new features appear in
this case, the most important of them being an interesting
broken-symmetry related to the interlayer biquadratic term.
\end{abstract}



\section{Introduction}
The study of the properties of magnetic multilayers has been one
of the most active fields in the last decade. The understanding of
a number of new and intriguing results, such as the biquadratic
exchange term in the free magnetic energy of the system, became an
exciting challenge from both theoretical and experimental point of
view. Until recently, it was found that the biquadratic exchange
coupling was too small when compared to the bilinear term, but
recent works have proved that it can play a remarkable role in the
properties of magnetic multilayers \cite{1,2,3,4}.

It is known that the magnetic properties can depend strongly on
the stacking pattern of the layers. In this respect, the physical
properties of a new class of artificial material, the so-called
quasiperiodic structures, became recently an attractive field of
research \cite{5}. Quasiperiodic structures, which can be
idealized as the experimental realization of a one-dimensional
quasicrystal, are composed from the superposition of two (or more)
building blocks that are arranged in a desired manner. They can be
defined as an intermediate state between an ordered system (a
periodic crystal) and a disordered one (an amorphous solid),
despite the purely deterministic rules used to generate them
\cite{6,7}. They also share a general property, which is perhaps
their most characteristic one, namely a complex fractal spectra of
energy, which can be considered as their basic
signature\cite{8,9,10,11,12}. Their first experimental realization
(in quasiperiodic GaAs-AlAs heterostructures) was carry out by
Merlin and collaborators \cite{13}, and since then they have
become a rapidly expanding object of theoretical and experimental
research.

Recently, Tsallis and collaborators \cite{14} presented a model,
based on the most well-known and simple deterministic fractal
geometry (the {\it  triadic Cantor set}), to study the
thermodynamic properties. This set is obtained through the
repetition of a simple rule: divide a given segment into three
equal parts, and then eliminate the central one. They showed that
the specific heat of such a system exhibits a very particular
behavior: it oscillates log-periodically around a mean value that
equals the fractal dimension of the spectrum. Connections with
natural spectra showing multifractal properties were also
demonstrated afterwards \cite{15,16}. Furthermore, in a recent
publication \cite{17}, we extended the model described by Tsallis
{\it et al}, to study the specific heat, of {\it real spin waves
modes} that propagate in quasiperiodic structures, with new
features of the specific heat behavior. Throughout these papers,
the classical Maxwell-Boltzmann statistics were used.

It is the aim of this work to investigate these properties even
further, by studying the influence of the biquadratic exchange
term in the specific heat spectra of spin waves in a magnetic
quasiperiodic structure obeying the Fibonacci sequence. Spin-wave
studies of magnetic systems, where there are contributions of
bilinear and biquadratic exchange interactions, are quite recent
\cite {18,19,20}. Such investigations renewed the interest in
spin-wave analysis, particularly for non-collinear configurations,
since it can give important information about the coupling
parameters \cite{21}. The biquadratic coupling in real systems
usually favors perpendicular alignment of the film magnetization,
whereas the bilinear one favors both parallel or antiparallel
configurations. The inclusion of both exchange terms leads to
interesting physical properties \cite{22}.

The plan of this work is as follows. In Section II we present our
general theoretical model, based on ferromagnetic films described
by the Heisenberg Hamiltonian with a biquadratic exchange term,
which are stacked following a Fibonacci sequence. The multifractal
spectra then obtained will be used to determine the specific heat
spectra for the Fibonacci quasiperiodic magnetic arrangement
described in Section III. The numerical results and the discussion
of their main features are presented in section IV.

\section{The model}
In this section we follow the spin-wave model of Ref.\ 20 for
quasiperiodic magnetic superlattices with biquadratic exchange
coupling. Specifically, we consider magnetic superlattices
composed of $n_A$ layers of material $A$ and $n_B$ layers of
material $B$. Materials $A$ and $B$ are simple cubic Heisenberg
ferromagnets with bulk bilinear exchange constants $J_A$ and
$J_B$, and lattice constant $a$. The spin quantum numbers of the
magnetic moments in each material are $S_A$ and $S_B$,
respectively. In our model we consider that within the bulk of
materials $A$ and $B$ the magnetic moments interact only by the
bilinear exchange couplings $J_A$ and $J_B$. However, across the
interfaces, the magnetic moments interact by interfacial bilinear
$(J_{bl}^I)$ and biquadratic $(J_{bq})$ exchange couplings. The
Hamiltonian for the bulk of each component is
\begin{equation}
{\mathcal H_{\alpha}}=(-1/2){\sum\limits_{i,j}{J_\alpha
\stackrel{\rightarrow }{S}_i\cdot \stackrel{\rightarrow
}{S}_j}}-g\mu _BH_0{\sum\limits_i{S_i^z}},
\end{equation}
where the sum in the first term is over nearest neighbors $j$,
$H_0$ is the applied magnetic field in the $z$-direction, and
$\alpha $ is equal to $A$ or $B$. Also, $g$ is the usual Land\'e
factor and $\mu_B$ is the Bohr magneton. The spins of sites that
are at the interfaces have a Hamiltonian which includes an
additional biquadratic exchange coupling term, namely:
\begin{equation}
{\mathcal H}_{bq}={\sum\limits_{i,j}{J_{bq} (\stackrel{\rightarrow
}{S}_i\cdot \stackrel{\rightarrow }{S}_j})^{2}}.
\end{equation}
It should be remarked that this term is responsible for the new
effects found in the specific heat reported in this paper.

The spin wave dispersion relation in a magnetic superlattice is
found by considering the wave solution inside each material and
applying appropriate boundary conditions at the interfaces. The
solutions for each material can be written as a linear combination
of the positive- and negative-going solutions of the bulk case,
i.e.,
\begin{eqnarray}
\lefteqn {S_i^{+}=\{ A_l\exp [i\stackrel{\rightarrow }{k}_A\cdot
(\stackrel{ \rightarrow }{r}-\stackrel{\rightarrow }{r}_{lA})]}
\nonumber \\ &&+A^\prime_l\exp [-i\stackrel{ \rightarrow
}{k}_A\cdot (\stackrel{\rightarrow }{r}-\stackrel{\rightarrow}{r
}_{lA})]\} \exp (-i\omega t)
\end{eqnarray}
\noindent in component A, cell {\it l}, and
\begin{eqnarray}
\lefteqn {S_i^{+}=\{ B_l\exp [i\stackrel{\rightarrow }{k}_B\cdot
(\stackrel{ \rightarrow }{r}-\stackrel{\rightarrow }{r}_{lB})]}
\nonumber \\ &&+B^\prime_l\exp [-i\stackrel{ \rightarrow
}{k}_B\cdot (\stackrel{\rightarrow }{r}-\stackrel{\rightarrow}{r
}_{lB})]\} \exp (-i\omega t)
\end{eqnarray}
\noindent in component B, cell {\it l}.

These solutions are linked together using the equation of motion
for the operator $S_i^+=S_i^x+iS_i^y$, which for a site $i$ at an
interface and after using the RPA approximation, is (see Ref.\ 20
for details)
\begin{eqnarray}
\lefteqn {\hbar \frac \partial {\partial t}S_i^{+}=g\mu
_BH_0S_i^{+}+\sum\limits_{n.n.}J_\alpha(S_jS_i^{+}-S_iS_j^{+})}
\nonumber \\ &&+\sum\limits_{n.n.}J_{bq}\{ \
[S_{i}^2(2S_j-1)-S_i(S_j-1)]S_j^{+}\nonumber
\\ &&-[S_{j}^2(2S_i-1)-S_j(S_i-1)]S_i^{+}\ \}.
\end{eqnarray}
The boundary conditions, after a tedious but straightforward
calculation, can be written in a matrix form like
\begin{equation}
{\mathbf M_{A}}\left[  \begin{array}{c}  A_l \\ \\ A^\prime_l
\\
\end{array}
 \right]
={\mathbf N_{B}}\left[  \begin{array}{c}  B_l \\ \\ B^\prime_l \\
\end{array}\right]
\end{equation}
and
\begin{equation}
{\mathbf M_{B}}\left[  \begin{array}{c}  B_l \\ \\ B^\prime_l
\\
\end{array}
 \right]
={\mathbf N_{A}}\left[  \begin{array}{c}  A_{l+1} \\ \\
A^\prime_{l+1}
\\
\end{array}\right].
\end{equation}
The explicit form of the above matrices can be found elsewhere
\cite{20}. Therefore, the explicit relation between the $lth$ and
$(l+1)th$ unit cell amplitudes is
\begin{equation}
\left[  \begin{array}{c}  A_{l+1} \\ \\ A^\prime_{l+1}
\\
\end{array}
 \right]
={ N^{-1}_{A}M_BN^{-1}_{B}M_A}\left[  \begin{array}{c} A_l
\\
\\ A^\prime_l
\\
\end{array}\right],
\end{equation}
Here $T=N_A^{-1}M_BN_B^{-1}M_A$ is a transfer matrix. This
equation, combined with the Bloch ansatz yields
\begin{equation}
 [{ T}-\exp(iQD)] \left[  \begin{array}{c} A_l
\\ \\ A^\prime_l
\\
\end{array}
 \right]
=0.
\end{equation}
The analogous relation between $(A_{l-1},A^\prime_{l-1})$ and
$(A_l,A^\prime_l)$ combined with equation (9) yields,
\begin{equation}
\cos (QD)=(1/2)Tr\left[ T\right].
\end{equation}
Here $Q$ is the Bloch wavevector of the collective excitation, and
$D$ is the size of the superlattice unit cell. Equation (10)
follows from the fact that $T$ is an unimodular 2x2 matrix, and it
describes the bulk modes of spin waves in a magnetic superlattice.
Once we know the form of the transfer matrix $T$, the bulk spin
wave spectrum is determined.

Let us briefly review the quasiperiodic sequence considered in
this work. First we recall the definition of a {\it substitution
sequence}. Take a finite set $\xi$ (here $\xi=\{A,B\}$) called an
{\it alphabet} and denote by $\xi^*$ the set of all words of
finite length that can be written in this alphabet. Now let
$\zeta$ be a map from $\xi$ to $\xi^*$ by specifying that $\zeta$
acts on a word by substituting each letter (e.g. A) of this word
by its corresponding image $\zeta(A)$. A sequence is then called a
{\it substitution sequence}, if it is a fixpoint of $\zeta$, i.e.
if it remains invariant when each letter in the sequence is
replaced by its image under $\zeta$. One of the most famous
substitution sequences is the so-called Fibonacci quasiperiodic
sequence, whose substitution rule is $A \rightarrow \zeta(A)=AB$,
$B \rightarrow \zeta(B)=A$. This sequence can also be constructed
by appending the $n-2th$ generation to the $n-1th$ one, i.e.,
$S_n=S_{n-1}S_{n-2}$ ($n\geq 2$). This algorithm requires the
initial conditions $S_0=B$ and $S_1=A$. Some of the first
Fibonacci generations are,
\begin{center}
$S_2=\left[A B \right]$  , $S_3=\left[ AB  A \right]$, $S_4=\left[
AB AA B \right]$,\ \ etc.
\end{center}
In a given generation $S_n$, the total number of building blocks
is given by the Fibonacci number $F_n$, which is obtained by the
relation $F_n=F_{n-1}+F_{n-2}$, with $F_0=F_1=1$. Also, $F_{n-1}$
and $F_{n-2}$ are the numbers of building blocks $A$ and $B$,
respectively. As the generation order increases ($n
>> 1$), the ratio $F_{n}/F_{n-1}$ approaches $\tau = (1+
\sqrt5)/2$, an irrational number which is known as the golden
mean. From an experimental perspective, a Fibonacci superlattice
is grown juxtaposing the two magnetic materials $A$ and $B$
according to the above described sequence of letters. It can also
be shown that the transfer matrices of consecutive Fibonacci
generations are related by a convenient recursive relation
\cite{9,20},
\begin{equation}
{T_{s_{n}}}= {T_{s_{n-2}}} \cdot {T_{s_{n-1}}}, \quad n\geq2.
\end{equation}
It is easy to see that, from the knowledge of the transfer
matrices $T_{S_{0}}$ and $T_{S_{1}}$, we can determine the
transfer matrix for any generation and consequently the spin-wave
dispersion relation. It should be remarked that the matrix
$T_{S_{2}}$ recovers the periodic case.

\section{Specific heat Spectra}
The spin wave fractal spectra for the  Fibonacci superlattices
with biquadratic exchange coupling is shown in Fig.\ 1 for a fixed
value of the in-plane dimensionless wavevector, namely
$k_{x}a=2.0$. We can clearly see the forbidden and allowed
energies versus the  Fibonacci generation number $n$, up to their
$10th$ generation, whose unit cell is composed of $F_9=55$ $A$'s
and $F_8=34$ $B$'s building blocks. The number of allowed bands is
equal to three times the Fibonacci number $F_{n}$ of the
correspondent generation. Notice that, as expected, for large $n$
the allowed band regions get narrower and narrower and they have a
typical Cantor set structure.

We address now  the specific heat of the spectra shown in Fig.\ 1.
The description below, which follows the lines of Ref.\ 17, is
general and has been successfully applied to many other banded
spectra (see also Refs.\ 23 and 24). In Fig.\ 1 each spectrum, for
a fixed generation number $n$, has $m$ allowed continuous bands.
We consider the level density within each band to be constant. The
partition function for the $nth$ generation is then given by:

\begin{equation}
Z_{n}=\int_{0}^{\infty}\rho (\epsilon)e^{-\beta
\epsilon}d{\epsilon},
\end{equation}
Here $\beta=1/T$ (by choosing the Boltzmann's constant $k_{B}=1$),
and we take the density of states $\rho (\epsilon)=1$. After a
straightforward calculation we can write $Z_{n}$ as

\begin{equation}
Z_{n}={\frac {1}{\beta}}\sum_{i=1,3,\ldots}^{2m-1}
e^{-\beta\epsilon_{i}}[1-e^{-\beta\Delta_{i}}].
\end{equation}
Here the subscript $n$ is the generation number, $m$ is the number
of allowed bands and $\Delta_{i}=\epsilon_{i+1}-\epsilon_{i}$ is
the difference between the top and bottom energy levels of each
band.

The specific heat is then given by
\begin{equation}
C_{n}(T)={\frac {\partial}{\partial T}}[T^2{\frac {\partial \ln
Z_{n}}{\partial T}}],
\end{equation}
which can be written as
\begin{equation}
C_{n}(T)=1+{\frac {\beta f_{n}}{Z_{n}}}-{\frac
{g^2_{n}}{Z^2_{n}}},
\end{equation}
with
\begin{equation}
f_{n}=\sum_{i=1,3,\ldots}^{2m-1}[ \epsilon ^2_{i}
e^{-\beta\epsilon_{i}}-\epsilon ^2_{i+1}e^{-\beta\epsilon_{i+1}}].
\end{equation}
and
\begin{equation}
g_{n}=\sum_{i=1,3,\ldots}^{2m-1}[ \epsilon_{i}
e^{-\beta\epsilon_{i}}-\epsilon_{i+1}e^{-\beta\epsilon_{i+1}}].
\end{equation}

Therefore, once we know the energy spectra of the spin waves which
propagates in a given sequence's generation of a quasiperiodic
structure, we can determine the associated specific heat by using
(15).
\section{Numerical results and discussions}

In this section we present numerical results obtained for the
specific heat of Fibonacci magnetic multilayers.  In our
calculations we have assumed the spin quantum numbers $S_A=1.0$
and $S_B=1.5$, and have taken values of the ratio between the
interlayer biquadratic and bilinear exchange terms
($R=J_{bq}/J_{bl}^I$) in accordance with experimental data
recently measured for different regions of the biquadratic and
bilinear exchange couplings' regime\cite{1,2,21}.

Fig.\ 2 shows the {\it log-log} plot of the spin wave specific
heat as a function of temperature in the low temperature regime
and in the absence of the biquadratic term ($R=0$).  We have taken
the dimensionless common in-plane wavevector to be $k_xa=2.0$. As
we can see,  there is an interesting harmonic oscillation of the
specific heat, which is a {\it log-periodic} function of the
temperature, i.e., $C_{n}(T)=\alpha C_{n}(aT)$, where $\alpha$ is
a constant, and $a$ an arbitrary number. These log-periodic
oscillations can be traced back to the log-periodicity of the
density of state's spectral staircase. They resemble very much the
shape of the devil's staircase obtained from idealized Cantor sets
\cite{15,16}. The number of oscillations depends on the generation
number of the Fibonacci sequence: the bigger the system, the
greater the number of oscillations. Most important, however, is
the well defined even and odd parity spectra related to the
generation number of the Fibonacci structure, with the amplitudes
of the latter being bigger than the amplitudes of the former. They
are due to the parity of the set of eigenvalues ($\epsilon_i$)
used to calculate the partition function given by (13). These
harmonic oscillations can be identified as the {\it signature} of
the Fibonacci structure, and it has no counterpart in the
idealized triadic Cantor set.

There is an important modification, however, when the biquadratic
coupling is present ($R\neq 0$). By contrast with the spectra
found in the previous work \cite {17} (but with other similarities
remaining), the two different symmetrical profiles of oscillations
are broken for different ratios between the biquadratic and
bilinear terms. To reinforce this intriguing behavior, we show, in
Figs.\ 3 and 4, the evolution of this broken-symmetry for $R=0.2$
(where the even-mode symmetry starts to be broken ) and $0.4$
(where the odd-mode symmetry also starts to be broken),
respectively. A possible explanation is that these different
behaviors arise because the biquadratic exchange term reduces the
effective coupling between the adjacent magnetic layers at the
interface, e.g., as may be deduced by examining the overall
coefficients of $S_i^{+}$ or $S_j^{+}$ in (5) (see also Ref.\ 20).
Also the biquadratic exchange induces long range correlations that
emphasize the quasiperiodicity of the system. In other words, it
looks like these effects make the whole structure {\it see better}
its quasiperiodicity, in effect increasing its degree of disorder
as $R$ increases! It turns out that the spin wave energy band
structure is more disordered and, as a consequence,  this is
reflected in the specific heat curves, which do not have the
well-defined standard profiles (actually the oscillations are
non-harmonic!) found  in the absence of the biquadratic term, as
well as presented for other excitations \cite {23,24}. This
argument is reinforced by previous works on the correlation
lengths of magnetic systems exhibiting biquadratic exchange
coupling (see, for example, S\o rensen and Young \cite{25}).

To summarize, we have proposed in this paper a realistic model to
study the specific heat contribution from the energy spectra of
{\it real} spin waves in quasicrystals of the Fibonacci type,
stressing the important role played by the biquadratic exchange
term. Certainly the theoretical predictions shown here can be
tested experimentally, and we encourage experimentalists  to carry
out such studies.

Further investigations for the case of the specific heat due to
propagation of mixed modes is currently under consideration, and
we hope to present these results in a later publication.

\vskip 0.5 cm

\noindent {\it Acknowledgements}: We would like to thank the
Brazilian Research Council CNPq and NSERC of Canada for financial
support. One of us (ELA) also thanks the hospitality of the
Department of Physics and Astronomy, University of Western
Ontario, where his contribution to this work was made.

\newpage
\centerline {\bf Figure Captions}

\begin{enumerate}

\item Energy spectra of spin waves for the quasiperiodic Fibonacci structure when a
biquadratic interlayer exchange term is present. Here we consider
the dimensionless in-plane wavevector $k_xa=2.0$.

\item  Log-log plot of the specific heat versus temperature for the generation numbers
of the Fibonacci quasiperiodic sequence in the absence of the
biquadractic term: (a) even generation numbers; (b) odd generation
numbers.

\item Same as Fig.\ 2, but for the ratio between the interlayer
biquadratic and bilinear exchange terms $R=0.2$. From this value
the even symmetry starts to be broken.

\item Same as Fig.\ 2, but for the ratio between the interlayer
biquadratic and bilinear exchange terms $R=0.4$. From this value
the odd symmetry also starts to be broken.

\end{enumerate}

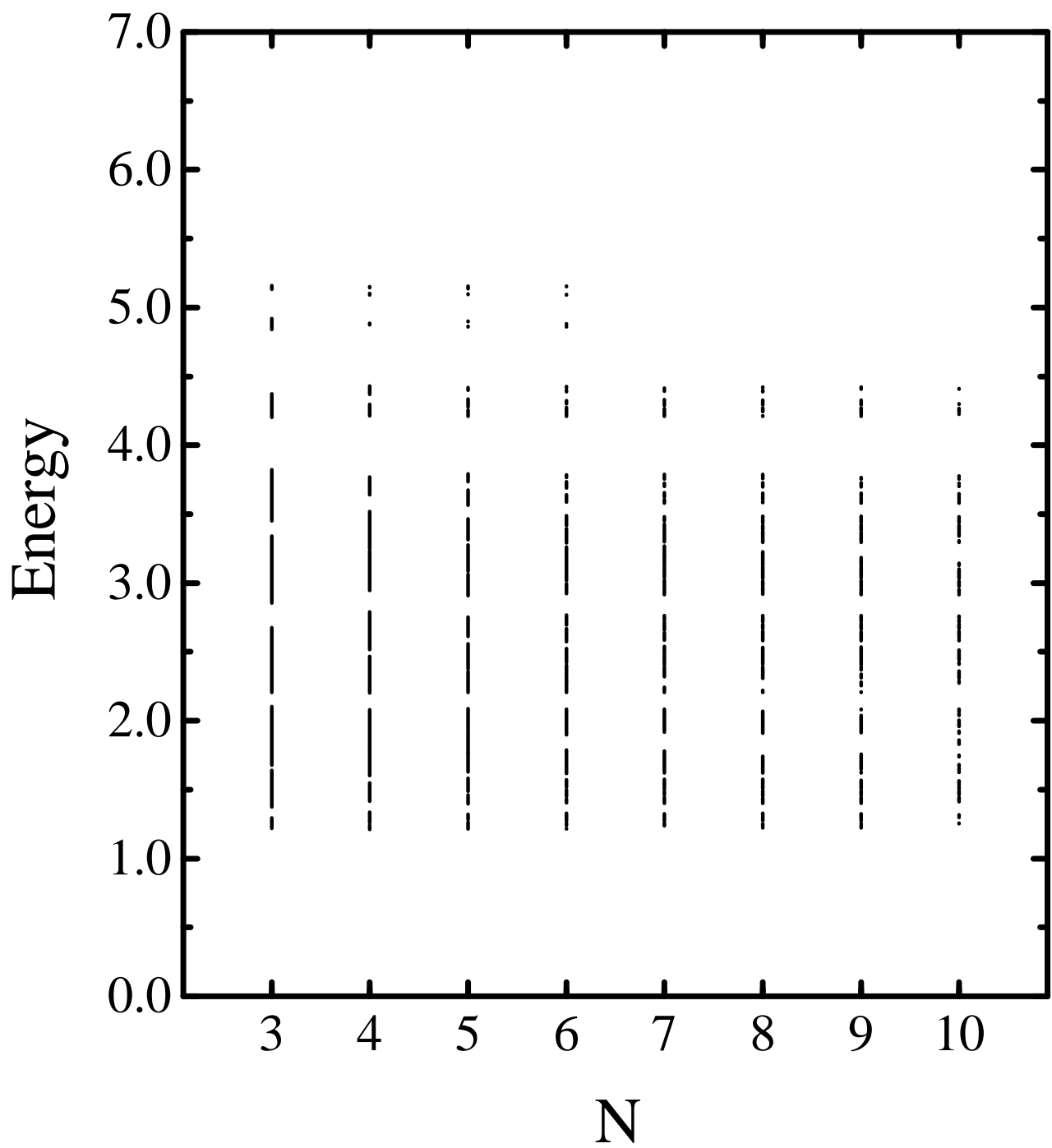
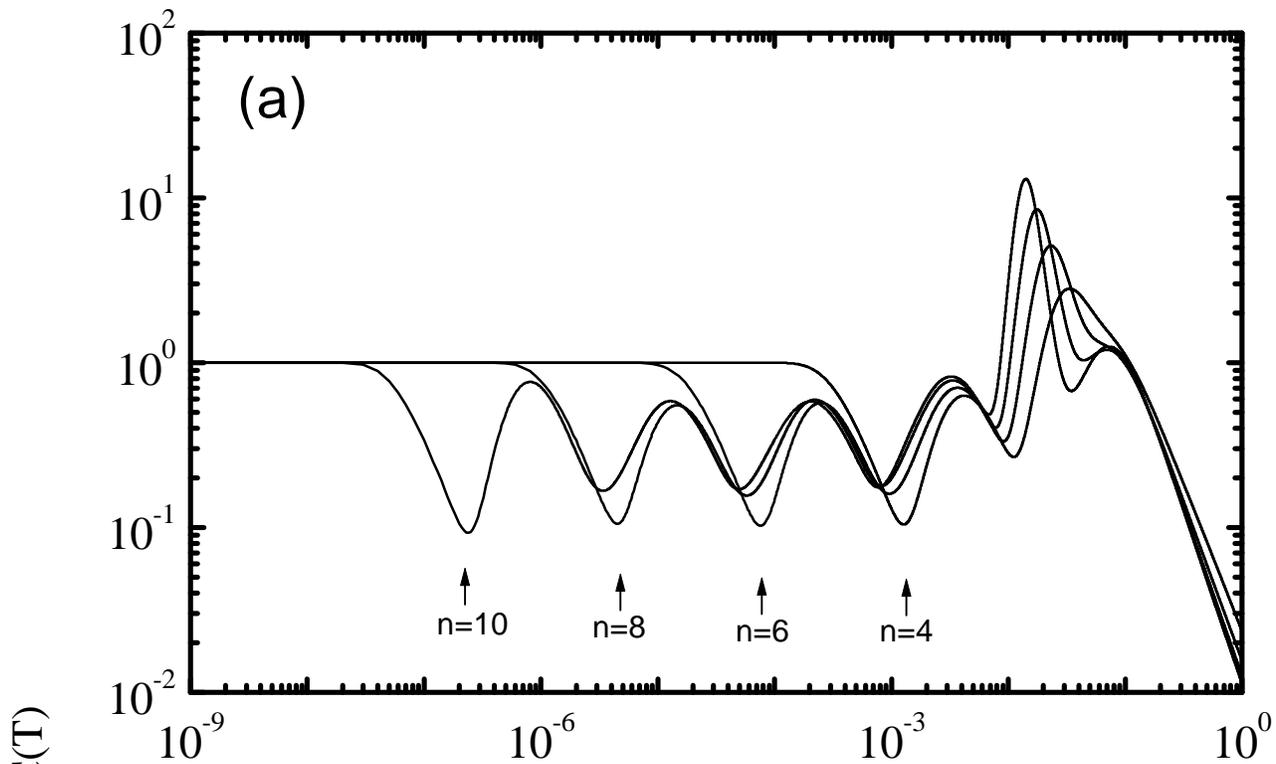
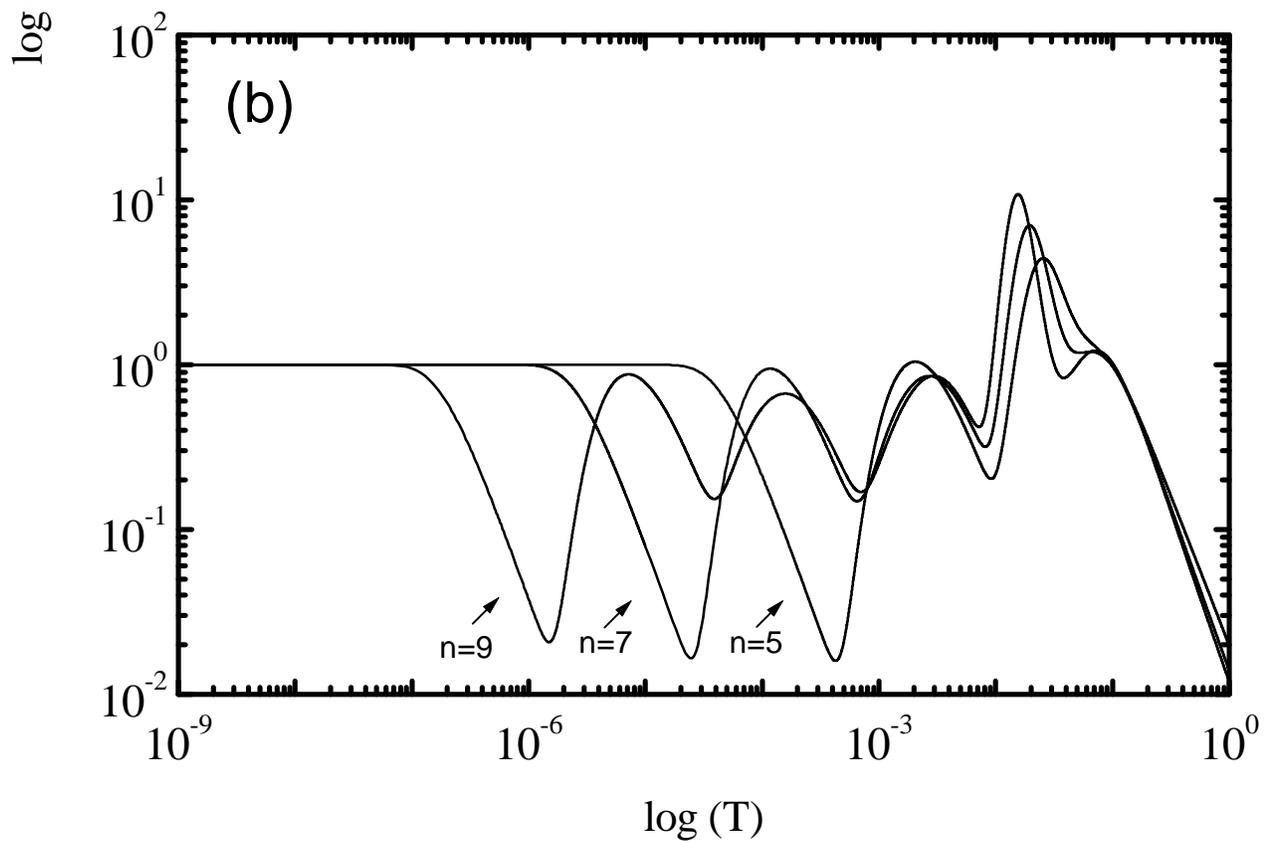
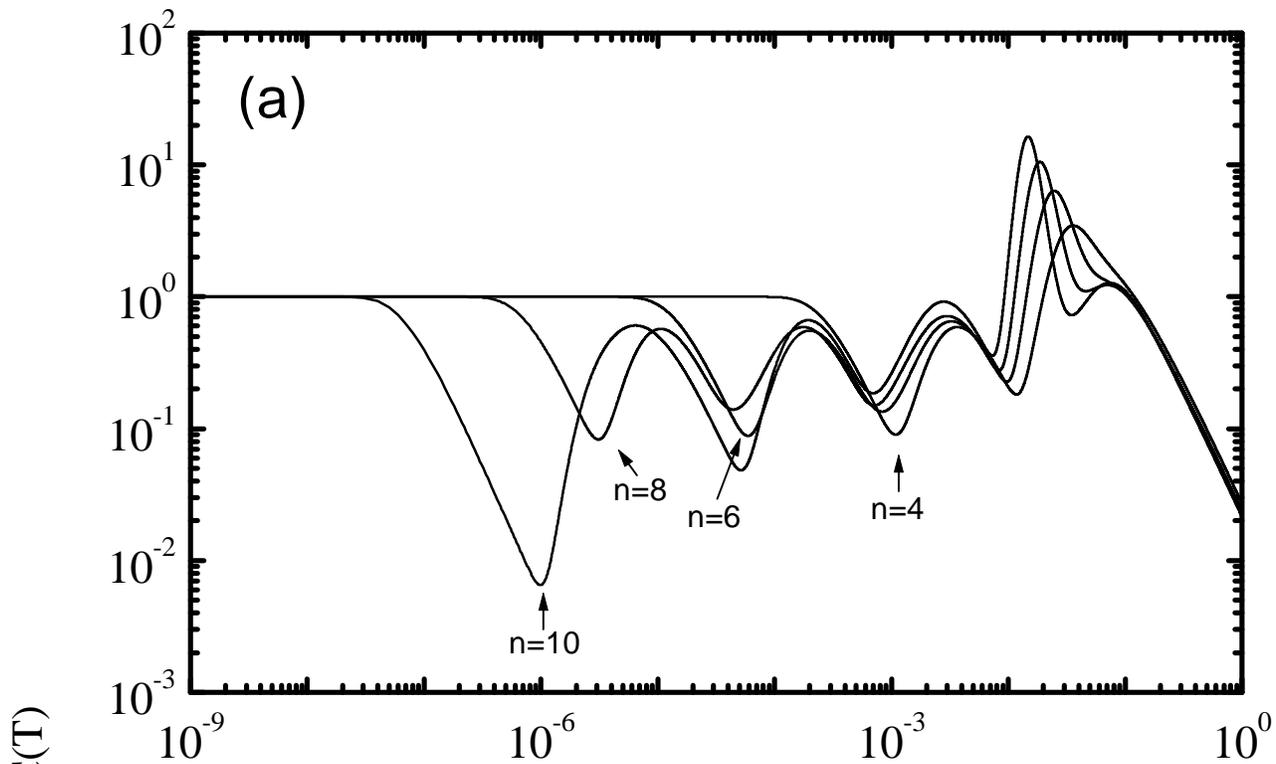
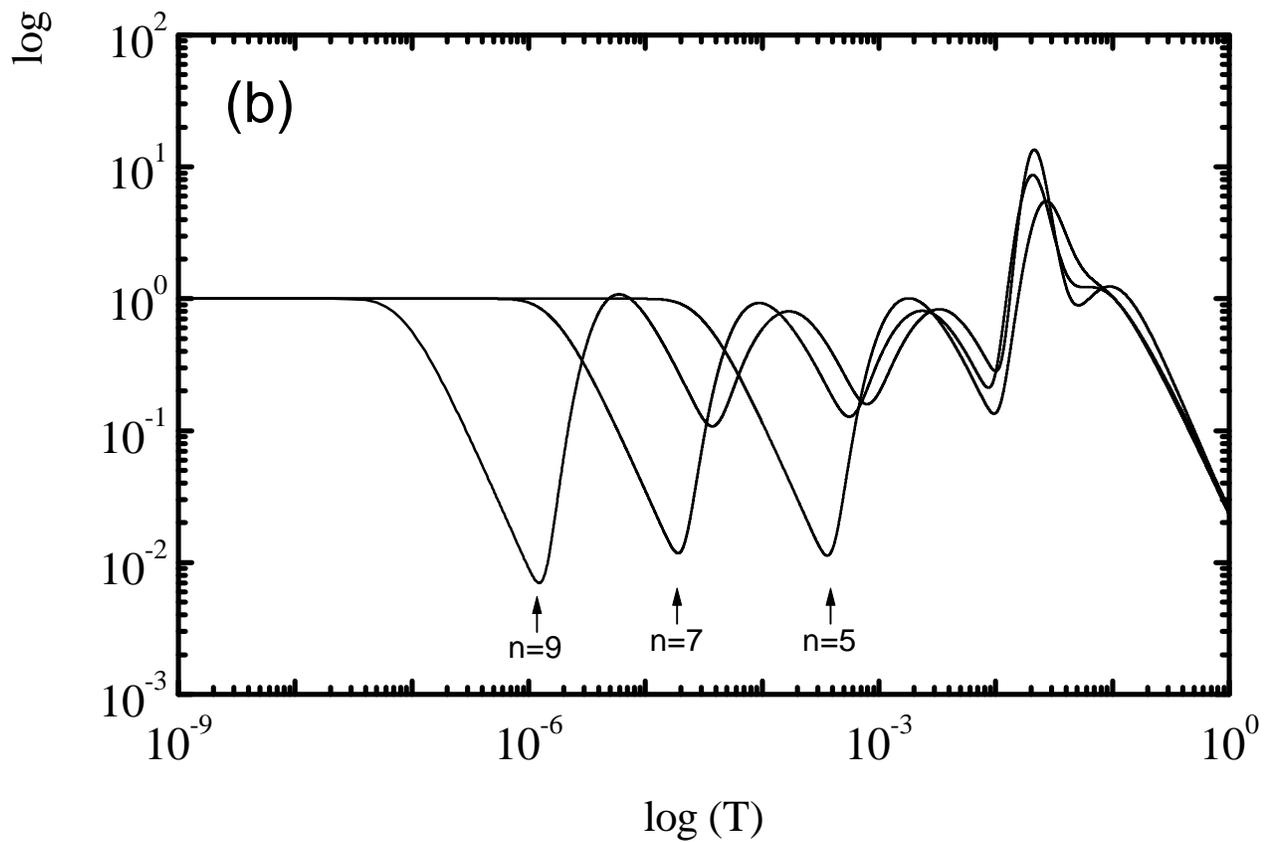
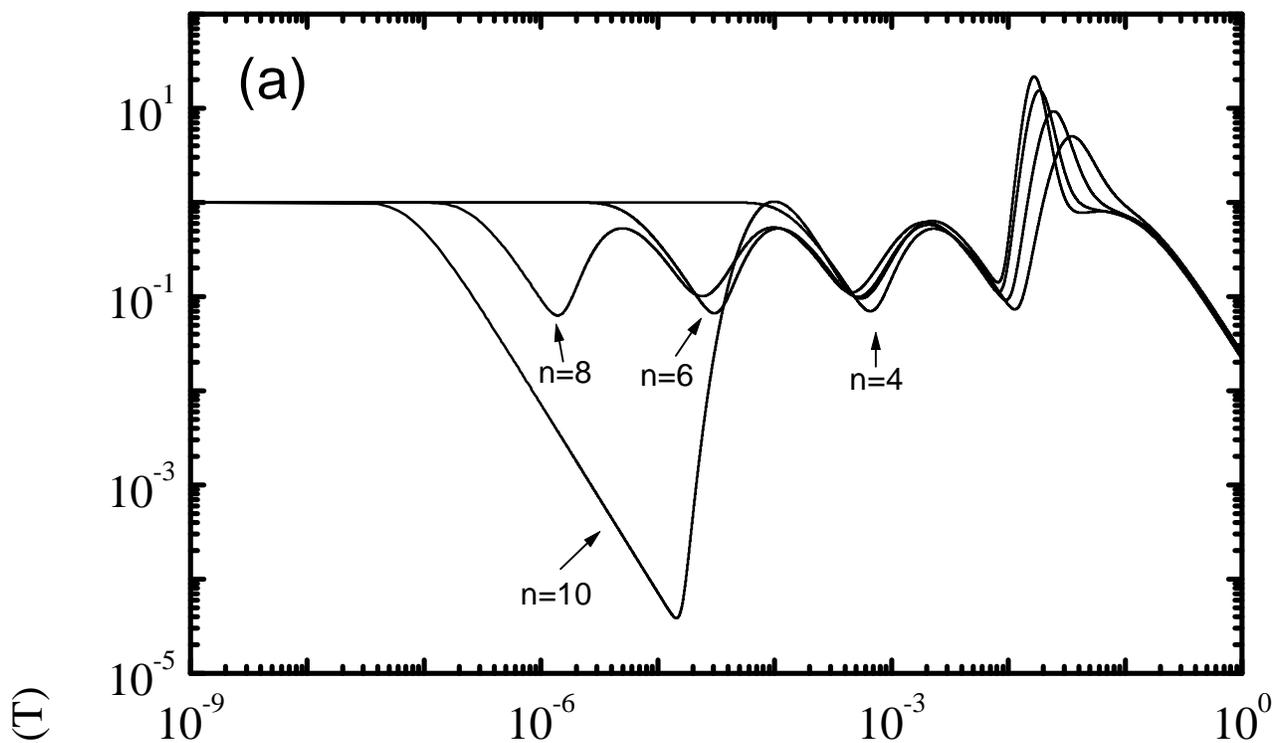
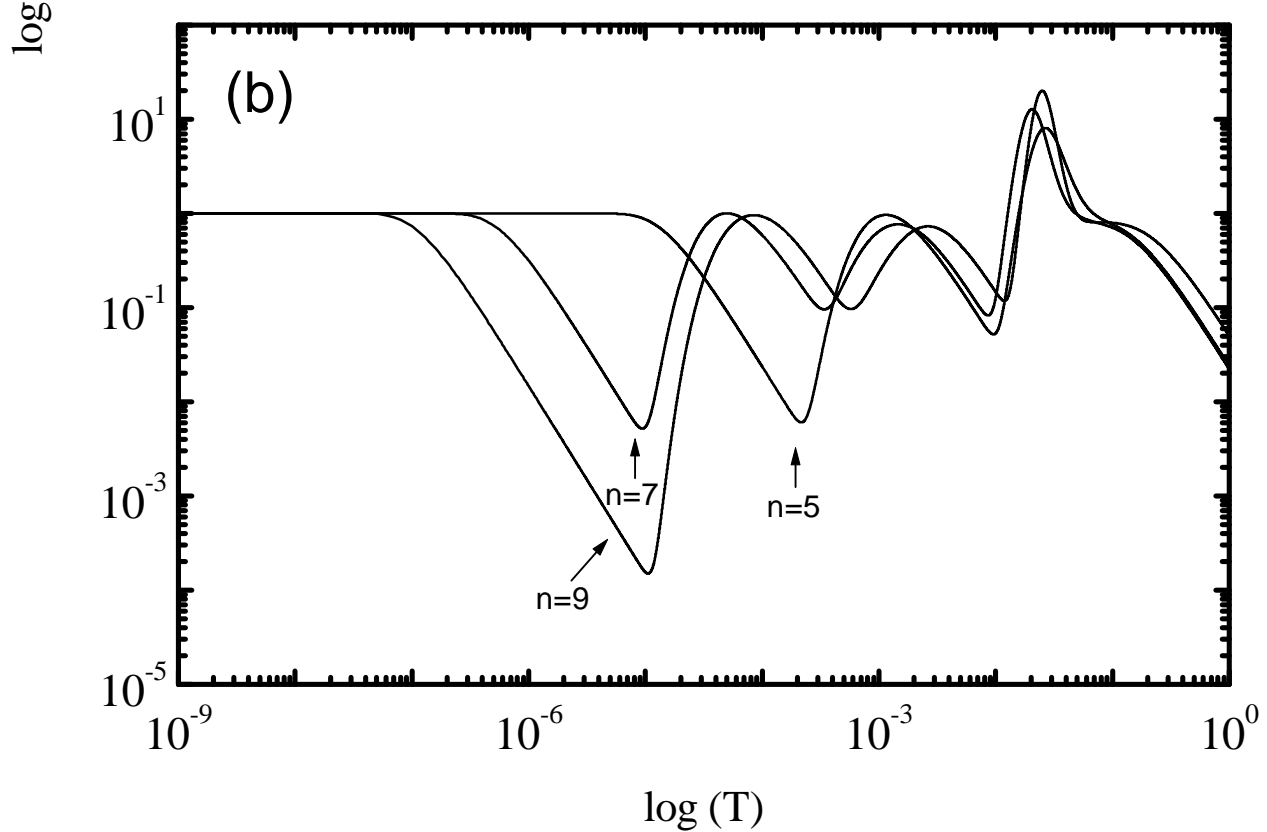
\end{document}